\documentclass[11pt]{iopart}

\usepackage{iopams} 
\usepackage{mathrsfs}
\usepackage{graphicx}
\usepackage{hyperref}
\usepackage{color}

\eqnobysec
\definecolor{CiteColor}{rgb}{0,0,0.35}
\definecolor{URLColor}{rgb}{0,0,0.35}
\hypersetup{colorlinks,citecolor=CiteColor,urlcolor=URLColor}

\newcommand{\ud}{\mathrm{d}}

\begin{document}

\title{Spacetime Symmetries and Kepler's Third Law}

\author{Alexandre Le Tiec}

\ead{letiec@umd.edu}

\address{Maryland Center for Fundamental Physics \& Joint Space-Science Institute, Department of Physics, University of Maryland, College Park, MD 20742, USA}

\begin{abstract}
The curved spacetime geometry of a system of two point masses moving on a circular orbit has a helical symmetry. We show how Kepler's third law for circular motion, and its generalization in post-Newtonian theory, can be recovered from a simple, covariant condition on the norm of the associated helical Killing vector field. This unusual derivation can be used to illustrate some concepts of prime importance in a general relativity course, including those of Killing field, covariance, coordinate dependence, and gravitational redshift.
\end{abstract}


\maketitle

\section{Introduction and summary}

One could hardly overstate the major role played by symmetries in physics. Symmetry considerations can very often drastically simplify the process of solving a given physics problem. At a more fundamental level, symmetries are deeply connected to the existence of conserved quantities (\textit{via} Noether's theorem), and they considerably restrain the span of admissible field theories in modern theoretical physics. Within Einstein's theory of general relativity (GR), which describes gravitation as a manifestation of the curvature of spacetime, the infinitesimal generators of isometries are called Killing vector fields. They are widely used in current research in gravitation theory, and are an essential part of one's education in relativistic gravity \cite{Car,Har,MTW,Schu,Wal,Wei}.

The existence of gravitational radiation is one of the most important predictions of GR. Observing gravitational waves would have a tremendous impact on physics, astrophysics, and cosmology \cite{SaSc.09}. A worldwide effort is currently underway to achieve the first direct detection, using kilometer-scale ground-based interferometers such as LIGO and Virgo, as well as future space-based antennas, such as the planned mission LISA. Binary systems composed of compact objects (neutron stars or black holes) are among the most promising sources of gravitational waves. However, the detection and analysis of these exceedingly weak signals require very accurate theoretical predictions, for use as template waveforms to be compared to the output of the detectors \cite{CuTh.02}. Hence, modeling the orbital dynamics and gravitational-wave emission of compact binary systems is a timely problem in relativistic astrophysics.

Except for the occurrence of a gradual inspiral driven by gravitational radiation-reaction, the orbital motion of stellar-mass compact-object binaries can be considered to be circular, to a very high degree of approximation.\footnote{However, there are some known astrophysical scenarii which may result in gravitational-wave signals from stellar-mass compact binaries with non-negligable eccentricities, such as binaries formed by capture in globular clusters, or having undergone the Kozai mechanism. Supermassive black hole binaries may also have significant eccentricities while entering the sensitivity band of future space-based detectors.} As long as the typical radiation-reaction timescale is much larger than the orbital period, \textit{i.e.} during most of the inspiral phase, the true orbital motion can be approximated by an adiabatic sequence of circular orbits. Mathematically, the approximation of an exactly closed circular orbit translates into the existence of a \textit{helical Killing vector} (HKV) field $K^\alpha$, along the orbits of which the spacetime geometry is invariant. In the full theory of GR, an exact helical symmetry requires incoming radiation to balance the outgoing radiation produced by the orbital motion \cite{GiSt.84,De.89,Kl.04}. Such unphysical incoming radiation can, however, be avoided by using various approximations to GR, such as the conformal flatness condition \cite{Go.al.02,Gr.al.02,Fr.al.02,Sh.al.04}, the post-Newtonian approximation in the conservative sector \cite{Bl.al.10,Bl.al2.10,Le.al.12}, or the extreme mass-ratio approximation at linear order \cite{De.08,Ke.al2.10}.

In this paper, we shall consider two non-spinning compact objects moving on exactly circular orbits. These will be modeled as point masses $m_A$ (with $A=1,2$), a prescription commonly adopted in the field of gravitational-wave source modeling (see, \textit{e.g.}, \cite{Bl.06,Po.al.11}). We will prove that the gradient $\nabla_\alpha$ of the norm $K^2$ of the helical Killing vector $K^\alpha$ must vanish along the worldlines of the particles:
\begin{equation}\label{geom}
	{\left( \nabla_\alpha K^2 \right)}_A = 0 \, .
\end{equation}
We then show how this simple, geometric result can be used to derive the main relation encoding the Newtonian orbital dynamics of the binary system, namely Kepler's third law (restricted to circular orbits), and its generalization in post-Newtonian theory. This unusual derivation can be used to illustrate numerous concepts of prime importance in a GR course, including those of Killing vector, covariance, coordinate dependence, and gravitational redshift.

This paper is organized as follows: In section \ref{sec:Killing} we summarize some well-known, yet useful properties of Killing vectors. We then discuss, in section \ref{sec:helical}, the physics of binary systems of point masses moving along circular orbits, introducing the notion of redshift observable in section \ref{subsec:redshift}, and proving the relation \eref{geom} in section \ref{subsec:proof}. The derivation of Kepler's third law, and its generalization in post-Newtonian theory, are discussed in sections \ref{subsec:KeplerNewt} and \ref{subsec:Kepler1PN}, respectively. Finally, section \ref{sec:K2_zA} is devoted to some further comments of physical relevance on the norm of the HKV and its link to the redshift observable.

\section{Some elementary properties of Killing vectors}\label{sec:Killing}

For the convenience of the reader, we start by summarizing a few elementary properties of Killing vector fields, which will be used extensively throughout this paper. The main property of a Killing field $k^\alpha$ is that it satisfies Killing's equation
\begin{equation}\label{Killing}
	\mathscr{L}_k g_{\alpha\beta} = \nabla_\alpha k_\beta + \nabla_\beta k_\alpha = 0 \, ,
\end{equation}
where $\nabla_\alpha$ is the covariant derivative compatible with the spacetime metric $g_{\alpha\beta}$, and $\mathscr{L}_k$ is the Lie derivative along $k^\alpha$. Equation \eref{Killing} expresses the invariance of the spacetime geometry along the integral curves of the Killing vector.

Furthermore, Killing vectors provide well-defined conserved quantities. Let $u^\alpha$ be the four-velocity of a test particle, tangent to its worldline, and normalized such that $g_{\alpha\beta} u^\alpha u^\beta = -1$. Then the scalar product $s \equiv k^\alpha u_\alpha$ is a constant of the motion along the timelike geodesic followed by the test mass:
\begin{equation}\label{cst_motion}
	\dot{s} \equiv u^\beta \nabla_\beta \left( u^\alpha k_\alpha \right) = \dot{u}^\alpha k_\alpha + u^\alpha u^\beta \nabla_\beta k_\alpha = 0 \, .
\end{equation}
This follows from the geodesic equations of motion, $\dot{u}^\alpha = 0$, and the antisymmetry of $\nabla_\alpha k_\beta$. Two familiar examples of such conserved quantities in curved spacetime are the energy per unit mass $e = - k_{(t)}^\alpha u_\alpha$ and angular momentum per unit mass $j = k_{(\varphi)}^\alpha u_\alpha$ of a test particle in orbit around a rotating black hole, where $k_{(t)}^\alpha$ and $k_{(\varphi)}^\alpha$ are Killing vectors associated with the stationarity and axisymmetry of the Kerr metric.

\begin{figure}[t!]
	\begin{center}
		\includegraphics[scale=0.61]{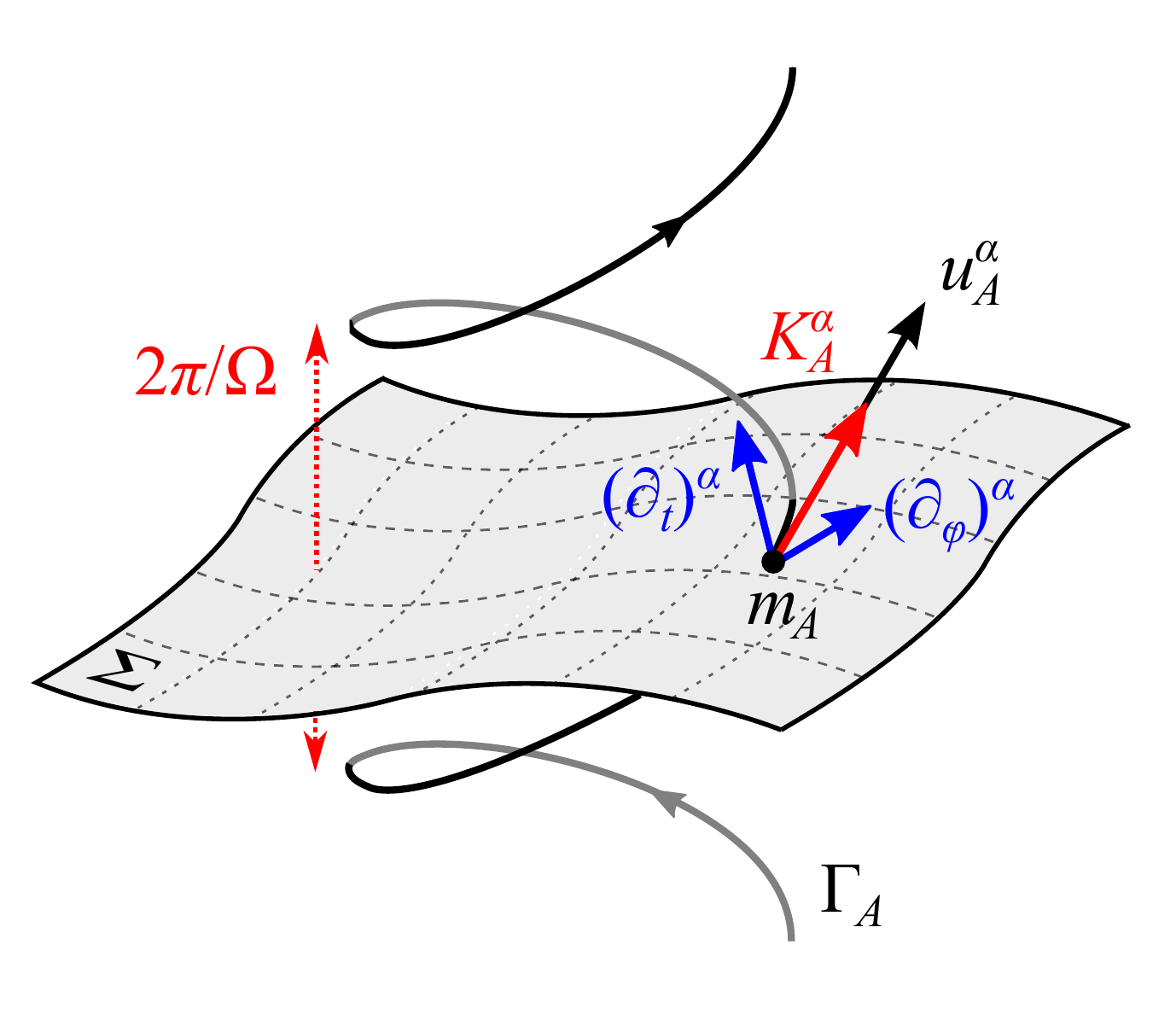}
		\caption{\footnotesize Spacetime diagram picturing a binary system of point masses $m_A$ ($A = 1,2$) on a circular orbit with constant azimuthal frequency $\Omega$. The helical Killing vector $K^\alpha = {(\partial_t)}^\alpha + \Omega \, {(\partial_\varphi)}^\alpha$ is aligned with the four-velocities $u_A^\alpha$ tangent to the worldlines $\Gamma_A$ of the particles. The redshift observables are given by the products $z_A = - K^\alpha_A u_\alpha^A$.}
		\label{fig:helical}
	\end{center}
\end{figure}

\section{Helically symmetric binary point-particle spacetimes}\label{sec:helical}

We now consider a binary system of non-spinning compact objects moving on a circular orbit. The neutron stars or black holes will be modeled as point particles with constant masses $m_A$ (with $A = 1,2$), and four-velocities $u_A^\alpha$ normalized to $g^A_{\alpha\beta} u_A^\alpha u_A^\beta = -1$. Note that the point masses $m_A$ are not test particles; their stress-energy tensor curves the geometry through Einstein's field equations. Their motion obeys the standard geodesic equations, $\dot{u}^\alpha_A = 0$, albeit in a \textit{regularized} metric $g^A_{\alpha\beta}$ such that the divergent self-fields of the point particles have been carefully subtracted \cite{Da.al.01,Ba.al2.02,DeWh.03,Bl.al.04}.

The spacetime metric of that binary system is neither stationary, nor axisymmetric; however it is invariant along the integral curves of a helical Killing vector $K^\alpha$. Far away from the binary, this field has the asymptotic behavior
\begin{equation}
	K^\alpha \to {(\partial_t)}^\alpha + \Omega  \, {(\partial_\varphi)}^\alpha \, ,
\end{equation}
where ${(\partial_t)}^\alpha$ and ${(\partial_\varphi)}^\alpha$ are part of the coordinate basis of an inertial frame of reference. The constant $\Omega$ is interpreted as the circular-orbit frequency of the binary system. Heuristically, $K^\alpha$ can be seen as the generator of time translations in a co-rotating frame. In particular, if you imagine yourself ``sitting'' on one of the particles, orbiting around the companion star, then you would observe no change in the local geometry. In other words, the metric is invariant along the worldlines of the particles: $\mathscr{L}_{u_A} g_{\alpha\beta} = 0$. This implies that the four-velocities of the particles must be aligned with the HKV evaluated at their respective coordinate locations (see figure \ref{fig:helical} for an illustration):
\begin{equation}\label{u_K}
	u^\alpha_A = u^T_A \, K^\alpha_A \quad \Longleftrightarrow \quad K^\alpha_A = z_A \, u^\alpha_A \, .
\end{equation}
As a coefficient of proportionality between two vectors, $u^T_A$, or equivalently $z_A \equiv 1 / u^T_A$, must be a scalar. It can be assigned several physical interpretations, which, in a GR course, would provide the opportunity to discuss the key notion of coordinate invariance.

\subsection{Redshift observable}\label{subsec:redshift}

\begin{figure}[t!]
	\begin{center}
		\includegraphics[scale=0.55]{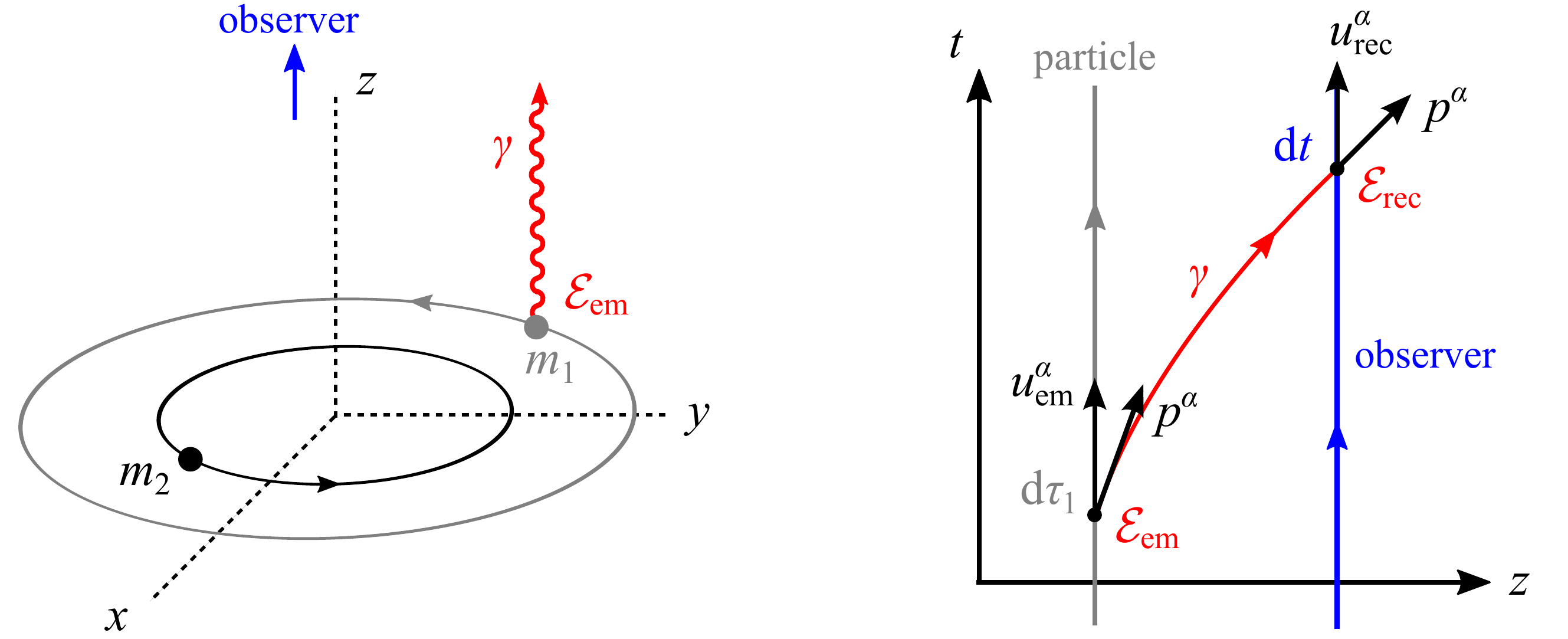}
		\caption{\footnotesize A photon $\gamma$ with four-momentum $p^\alpha$ is emitted from the particle $m_1$ with four-velocity $u_\mathrm{em}^\alpha = u_1^\alpha$, and received far away from the binary system, by a distant inertial observer with four-velocity $u_\mathrm{rec}^\alpha = {(\partial_t)}^\alpha$, along the helical symmetry axis $z$ perpendicular to the orbital plane. Detweiler's observable $z_1$ measures the redshift $\mathcal{E}_\mathrm{rec} / \mathcal{E}_\mathrm{em}$ of the photon, or equivalently the ratio $\ud \tau_1 / \ud t$.}
		\label{fig:redshift}
	\end{center}
\end{figure}

First, contracting the relation \eref{u_K} with the four-velocity $u_\alpha^A$, and remembering \eref{cst_motion}, we notice that $z_A$ is a constant of the motion associated with the helical symmetry:
\begin{equation}
	z_A = - K^\alpha_A u_\alpha^A = \mathrm{const.}
\end{equation}
Furthermore, it can easily be shown that $z_A$ measures the redshift of light rays emitted from particle $A$, and received far away from the binary, along the helical symmetry axis perpendicular to the orbital plane \cite{De.08}: Let $p^\alpha$ be a four-vector tangent to the worldline of such a light ray, \textit{e.g.} the four-momentum of the associated ``photon''. Then the ratio of the photon energy at reception and emission is given by
\begin{equation}
	\frac{\mathcal{E}_\mathrm{rec}}{\mathcal{E}_\mathrm{em}} = \frac{{(u^\alpha p_\alpha)}_\mathrm{rec}}{{(u^\alpha p_\alpha)}_\mathrm{em}} = \frac{{(K^\alpha p_\alpha)}_\mathrm{rec}}{u_A^T{(K^\alpha p_\alpha)}_\mathrm{em}} = z_A \, .
\end{equation}
We made use of the equality $u^\alpha_\mathrm{rec} = K^\alpha_\mathrm{rec} = {(\partial_t)}^\alpha$ between the four-velocity of the observer and the HKV (since ${(\partial_\varphi)}^\alpha = 0$ along the helical symmetry axis), of the relationship \eref{u_K} at the location of the emitter, and of the conservation of $K^\alpha p_\alpha$ along the null geodesic followed by the photon. (The proof of that last point is identical to that given in \eref{cst_motion}, with the substitutions $u^\alpha \longrightarrow p^\alpha$ and $k^\alpha \longrightarrow K^\alpha$.) See figure \ref{fig:redshift} for an illustration of this Gedankenexperiment. Following Detweiler \cite{De.08}, we shall thus refer to $z_A$ as the ``redshift observable'' of particle $A$.

In addition, in a cylindrical coordinate system $\{ ct, \rho, \varphi, z \}$ adapted to the helical symmetry, \textit{i.e.} such that the expression $K^\alpha = {(\partial_t)}^\alpha + \Omega \, {(\partial_\varphi)}^\alpha$ holds everywhere,\footnote{In such an adapted coordinate system, any scalar (or component of a tensor) $F$ that respects the helical symmetry must satisfy $\mathscr{L}_K F = \left( \partial_t + \Omega \, \partial_\varphi \right) F = 0$, and thus depends on the coordinate time $t$ and the azimuthal angle $\varphi$ only in the combination $\varphi - \Omega t$.} and not merely far away from the binary, Eq.~\eref{u_K} implies
\begin{equation}
	z_A = \left( u^0_A \right)^{-1} = \frac{\ud \tau_A}{\ud t} \, .
\end{equation}
Hence the redshift observable coincides with the inverse time component of the four-velocity of the particle, or equivalently with the ratio of the proper times elapsed along the worldlines of the particle and of the distant inertial observer (\textit{cf.} figure \ref{fig:redshift}). This last interpretation is in agreement with the usual notion of redshift (gravitational redshift and/or Doppler effect).

\subsection{Geometric characterization of the binary dynamics}\label{subsec:proof}

We now have at our disposal all of the concepts and results necessary to prove \eref{geom}, namely that the spacetime gradient of the norm $K^2 \equiv g_{\alpha\beta} K^\alpha K^\beta$ of the Killing field must vanish at the location of each particle. The derivation goes as follows:
\begin{equation}
	\hspace{-1.2cm} \frac{1}{2} {\left( \nabla_\alpha K^2 \right)}_A = K_A^\beta {\left( \nabla_\alpha K_\beta \right)}_A = - z_A u_A^\beta {\left( \nabla_\beta K_\alpha \right)}_A = - z_A \left( \dot{z}_A u^A_\alpha + z_A \dot{u}^A_\alpha \right) = 0 \, .
\end{equation}
We successively made use of the relationship \eref{u_K} between $K^\alpha_A$ and $u^\alpha_A$, of Killing's equation, of the geodesic equations of motion, and of the fact that $z_A$ is a constant of the motion. (The covariant derivative ${(\nabla_\beta K_\alpha)}_A$ being evaluated along the four-velocity $u_A^\beta$, we can replace $K_\alpha$ by its value along the worldline of particle $A$, namely $K^A_\alpha = z_A u^A_\alpha$.) Although pretty straightforward, that proof would provide a good exercise for students.

The formula \eref{geom} is simple and elegant; it is covariant, and only makes reference to well-defined geometrical concepts in GR. It implies, in particular, that $K^2(\mathbf{x})$ has extrema at the coordinate locations of the point particles (see figure \ref{fig:K}). Furthermore, we shall now show that \eref{geom} encodes a well-known result of classical mechanics, namely Kepler's third law (for circular orbits), and its generalization in post-Newtonian theory.

\section{Kepler's third law and its general relativistic generalization}\label{sec:Kepler}

\subsection{Newtonian gravity}\label{subsec:KeplerNewt}

We consider first the Newtonian approximation of the full theory of general relativity, \textit{i.e.} the leading-order results in the formal limit $c^{-1} \to 0$. In that weak-field, small velocity approximation, the spacetime metric expressed in cylindrical coordinates $\{ ct, \rho, \varphi, z \}$ takes the form
\begin{equation}\label{metric}
	\ud s^2 = \left( -1 + \frac{2U}{c^2} \right) c^2 \ud t^2 + \ud \rho^2 + \rho^2 \ud \varphi^2 + \ud z^2 \, ,
\end{equation}
where the Newtonian gravitational potential reads $U(t,\mathbf{x}) = \sum_A G m_A / \vert \mathbf{x} - \mathbf{y}_A(t) \vert$, with $\mathbf{y}_A(t)$ the coordinate trajectory of the mass $m_A$. In our adapted coordinates such that $K^\alpha = {(\partial_t)}^\alpha + \Omega \, {(\partial_\varphi)}^\alpha$, the norm of the Killing field reads $K^2 = g_{00} + 2 \Omega g_{0\varphi} / c + \Omega^2 g_{\varphi\varphi} / c^2$. With the explicit expression \eref{metric} of the metric, this gives
\begin{equation}\label{K2}
	K^2 = -1 + \frac{2U}{c^2} + \frac{\rho^2 \Omega^2}{c^2} \, .
\end{equation}

In order to apply the result \eref{geom}, we need first to compute the partial derivatives of $K^2$. Using the Cartesian coordinates $\{ x^i \}$ associated with the cylindrical coordinates $\{ \rho, \varphi, z \}$ in the usual way, namely $x^1 = \rho \cos{\varphi}$, $x^2 = \rho \sin{\varphi}$, and $x^3 = z$, we have
\begin{equation}
	\partial_t K^2 = \frac{2}{c^2} \, \partial_t U \quad \mathrm{and} \quad \partial_i K^2 = \frac{2}{c^2} \left( \partial_i U + \rho \, \Omega^2 n^i \right) ,
\end{equation}
where $\mathbf{n} = (\cos{\varphi},\sin{\varphi},0)$ is the unit vector in the orbital plane $z = 0$. Focusing first on the spatial components, the relationship \eref{geom} implies
\begin{equation}
	- \rho_A \Omega^2 n_A^i = {(\partial_i U)}_A \, .
\end{equation}
We recognize Newton's second law expressing the equality of the centripetal acceleration of body $A$ and of the Newtonian gravitational force exerted on the body $A$ by the body $B \neq A$. Computing the forces explicitely, we find\footnote{The singular self-forces of the point masses ought to be subtracted. This can be done by means of a suitable regularization method, such as dimensional regularization.}
\begin{equation}\label{Newton}
	\rho_1 \Omega^2 = \frac{G m_2}{r^2} \quad \mathrm{and} \quad \rho_2 \Omega^2 = \frac{G m_1}{r^2} \, ,
\end{equation}
where $r \equiv \vert \mathbf{y}_1 - \mathbf{y}_2 \vert = \rho_1 + \rho_2$ is the coordinate separation between the two point particles. We may then add up equations \eref{Newton}, or remember that in the center-of-mass frame $\rho_1 = r \, m_2 / m$ and $\rho_2 = r \, m_1 / m$, with $m = m_1 + m_2$ the total mass of the binary. Both solutions yield Kepler's third law
\begin{equation}\label{Kepler}
	\Omega^2 = \frac{Gm}{r^3} \, ,
\end{equation}
which is recovered here in the particular case of circular motion. (Our derivation cannot be extended to generic eccentric orbits, for which the helical symmetry is lost.) On the other hand, the time component of \eref{geom} is identically satisfied, as ${(\partial_t U)}_A \propto \mathbf{v}_B \cdot \mathbf{n}_{12}$ vanishes for circular orbits, $\mathbf{v}_B = \ud \mathbf{y}_B / \ud t$ being the coordinate velocity of the particle $B \neq A$, and $\mathbf{n}_{12}$ the unit vector pointing from $m_2$ to $m_1$. 

In the Newtonian limit, the geometric condition \eref{geom} can be related to a standard result of classical mechanics: It is well-known that in a frame rotating at the angular rate $\Omega$ with respect to a mass-centered, inertial  frame of reference, both point masses can be described as being at rest, sitting at local minima of the effective potential
\begin{equation}\label{Ueff}
	U_\mathrm{eff} \equiv U + \frac{1}{2} \rho^2 \Omega^2 \, ,
\end{equation}
the repulsive centrifugal force deriving from the harmonic potential $\frac{1}{2} \rho^2 \Omega^2$ balancing exactly the attractive gravitational force deriving from the Newtonian potential $U$. Now, comparing \eref{K2} and \eref{Ueff}, we note that $U_\mathrm{eff} \propto K^2 + 1$; hence the condition ${(\partial_i K^2)}_A = 0$ from which we derived Kepler's third law \eref{Kepler} is nothing but the result ${(\partial_i U_\mathrm{eff})}_A = 0$ in disguise. See figure \ref{fig:K} for an illustration. The scalar field $K^2$ can thus be thought of as a relativistic generalization of the effective potential $U_\mathrm{eff}$, and the condition \eref{geom} as a covariant generalization of the Newtonian result ${(\partial_i U_\mathrm{eff})}_A = 0$.

\subsection{Post-Newtonian gravity}\label{subsec:Kepler1PN}

Since \eref{geom} is valid beyond the Newtonian limit, the relation \eref{Kepler} can be extended to include corrections coming from the full theory, by keeping known post-Newtonian terms in the metric $g_{\alpha\beta}$. For instance, at first post-Newtonian (1PN) order, \textit{i.e.} including the relativistic corrections $\mathcal{O}(c^{-2})$ to the Newtonian expression \eref{metric}, the metric reads (in Cartesian-like harmonic coordinates) \cite{Bl.al.98}
\numparts
	\begin{eqnarray}
		g_{00} &= -1 + \frac{2 G m_1}{c^2 r_1} + \frac{1}{c^4} \left[ \frac{G m_1}{r_1} \left( 4 \mathbf{v}_1 \cdot \mathbf{v}_1 - (\mathbf{n}_1 \cdot \mathbf{v}_1)^2 \right) - \frac{2 G^2 m_1^2}{r_1^2} \right. \nonumber \\ &\left. \qquad\quad\!\! + \, G^2 m_1 m_2 \left( - \frac{2}{r_1 r_2} - \frac{r_1}{2 r_{12}^3} + \frac{r_1^2}{2 r_2 r_{12}^3} - \frac{5}{2 r_2 r_{12}} \right) \right] \nonumber \\ &\qquad\quad\!\! + (1 \leftrightarrow 2) + \mathcal{O}(c^{-6}) \, , \label{g00} \\
		g_{0i} &= - \frac{4 G m_1}{c^3 r_1} v_1^i - \frac{4 G m_2}{c^3 r_2} v_2^i + \mathcal{O}(c^{-5}) \, , \\
		g_{ij} &= \delta_{ij} \left( 1 + \frac{2 G m_1}{c^2 r_1} + \frac{2 G m_2}{c^2 r_2} \right) + \mathcal{O}(c^{-4}) \, , \label{gij}
	\end{eqnarray}
\endnumparts
where $\delta_{ij}$ is the Kronecker symbol, $r_{12} \equiv r = \vert \mathbf{y}_1 - \mathbf{y}_2 \vert$ and $r_A = \vert \mathbf{x} - \mathbf{y}_A \vert$ are defined in terms of the Euclidean norm, and $\mathbf{n}_A = (\mathbf{x} - \mathbf{y}_A) / r_A$. Transforming the metric \eref{g00}--\eref{gij} to cylindrical coordinates, and repeating the calculation detailed in section \ref{subsec:KeplerNewt}, we recover from ${(\partial_i K^2)}_A = 0$ the known generalization of Kepler's third law at 1PN order (in harmonic coordinates), namely \cite{Bl.al.98}
\begin{equation}\label{Kepler1PN}
	\Omega^2 = \frac{Gm}{r^3} \left\{ 1 + \left( -3 + \nu \right) \frac{Gm}{c^2r} + \mathcal{O}(c^{-4}) \right\} .
\end{equation}
At that order of approximation, the calculation involves a crucial contribution coming from the general relativistic frame-dragging effect, through the metric component $g_{0\varphi}$. The 1PN coefficient $(-3+\nu)$ in \eref{Kepler1PN} involves the symmetric mass ratio $\nu \equiv m_1 m_2 / m^2$, such that $\nu = 1/4$ for an equal-mass binary, and $\nu \to 0$ in the extreme mass-ratio limit. (The relation ${(\partial_t K^2)}_A = 0$ is still found to be satisfied identically.)

In a general relativity course, this derivation would provide the occasion to discuss the key notions of covariance and coordinate dependence using a familiar example: since \eref{geom} is covariant, it conveys a physically meaningful result, independent of a particular choice of coordinates. By contrast, the generalized version \eref{Kepler1PN} of Kepler's third law is coordinate-dependent; the 1PN coefficient could be different from $(-3+\nu)$ if the relationship between the invariant frequency $\Omega$ and the coordinate-dependent separation $r$ was expressed in another coordinate system. But the precise way in which the function $\Omega(r)$ changes, depending on the coordinate system used to write the post-Newtonian metric, is precisely encoded in the covariance of \eref{geom}.

While the relation between $\Omega$ and $r$ is coordinate-dependent, the functions $z_A(\Omega)$ are coordinate-invariant; they have recently been computed up to very high orders in the post-Newtonian approximation \cite{De.08,Bl.al.10,Bl.al2.10,Le.al.12}. For example, the 1PN-accurate result for the redshift of particle $1$ reads (we assume $m_1 \leqslant m_2$)
\begin{eqnarray}\label{z1}
	z_1 &= 1 + \left( - \frac{3}{4} - \frac{3}{4} \sqrt{1-4\nu} + \frac{\nu}{2} \right) x \nonumber \\ &+ \left( - \frac{9}{16} - \frac{9}{16} \sqrt{1-4\nu} - \frac{\nu}{2} - \frac{\nu}{8} \sqrt{1-4\nu} + \frac{5}{24} \nu^2 \right) x^2 + \mathcal{O}(x^3) \, ,
\end{eqnarray}
where $x \equiv (G m \Omega / c^3)^{2/3}$ is a dimensionless, post-Newtonian parameter $\mathcal{O}(c^{-2})$ related to the circular-orbit frequency $\Omega$ that can be measured by a distant observer.

\section{Helical Killing vector and redshift}\label{sec:K2_zA}

From the Newtonian result \eref{K2}, we have the asymptotic behavior $K^2 \sim (\rho \, \Omega / c)^2 > 0$ in the limit $\rho \to +\infty$, which indicates that the HKV is spacelike far away from the helical symmetry axis. Close to the binary system, however, and along the worldlines of the particles in particular, the HKV is timelike ($K^2 < 0$) [remember \eref{u_K}]. In between, there must exist a worldtube over which the HKV is null ($K^2 = 0)$. This hypersurface is usually referred to as the ``light cylinder''. In the flat spacetime limit $m_A \to 0$, its radius is simply $\rho = c / \Omega$. (In the language of classical mechanics, this is the distance from the axis of rotation for which the velocity of an observer rotating at the angular rate $\Omega$ reaches the vacuum speed of light $c$.)

Furthermore, when evaluated at the coordinate locations of the particles themselves, the norm of the HKV is directly related to the redshift observables. Indeed, from \eref{u_K} we immediately get $z_A^2 = - K_A^2$. In the Newtonian limit, we thus have\footnote{Replacing $U_1$ and $\rho_1$ by their known Newtonian expressions in terms of the frequency $\Omega$, one can easily recover from Eq.~\eref{zA2} the Newtonian contribution to the invariant relation \eref{z1}.}
\begin{equation}\label{zA2}
	z_A^2 = 1 - \frac{2 U_A}{c^2} - \frac{\rho_A^2 \Omega^2}{c^2} \, .
\end{equation}
This result is consistent with the interpretation of $z_A$ as a measure of the redshift of light rays, as discussed in section \ref{subsec:redshift}. The observable $z_A$ has two contributions: (i) a term proportional to the Newtonian potential $U_A$ evaluated at the coordinate location of particle $A$, which gives the gravitational redshift, or Einstein effect, and (ii) a term involving the relative velocity $v_A = \rho_A \Omega$ with respect to the distant observer, yielding a transverse Doppler effect. (In the flat spacetime limit $m_A \to 0$, we recover the special relativistic result $z_A = \sqrt{1 - v_A^2/c^2}$.)

\begin{figure}[t!]
	\begin{center}
		\includegraphics[scale=0.35]{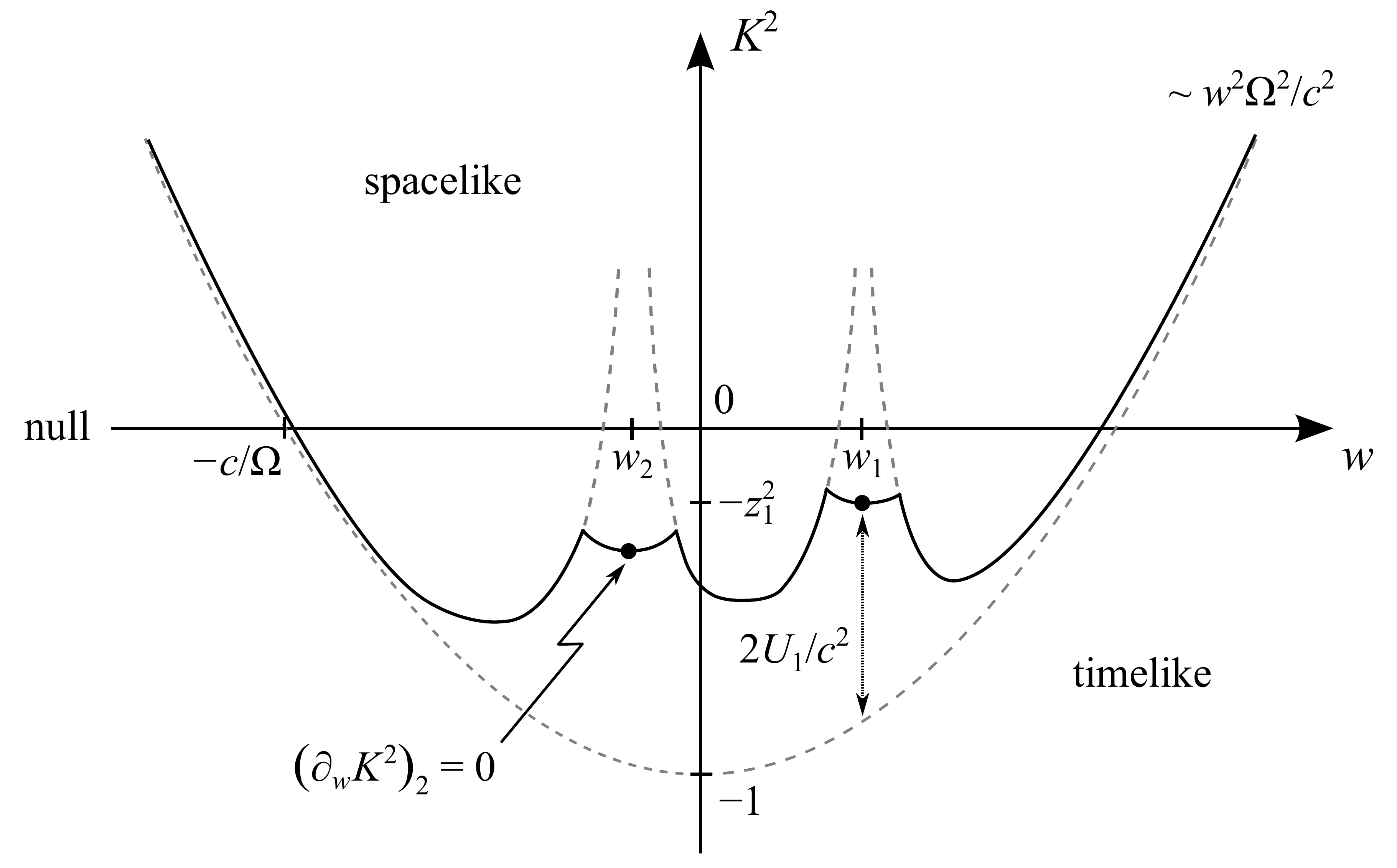}
		\caption{\footnotesize The norm $K^2$ of the helical Killing field $K^\alpha$ as a function of the coordinate $w$ along the direction joining the two particles, within the orbital plane.}
		\label{fig:K}
	\end{center}
\end{figure}

Based on the previous discussion, the function $K^2(w)$ is depicted schematically in figure \ref{fig:K}, with $w = \rho \cos{(\varphi - \varphi_1)}$ the coordinate along the direction joining the two particles, within the orbital plane. (The divergent Newtonian self-fields of the two point masses are shown in dashed lines; these are well-known artifacts of the use of point particles to model the actual physical compact stars, which are extended objects.) A more quantitative analysis of the function $K^2(\mathbf{x})$ in Newtonian (or post-Newtonian) gravity, in and out of the orbital plane, could be a useful exercise for students. They could, for instance, be asked to plot that function for different values of $\{\Omega, m_1, m_2\}$.

\ack
It is a pleasure to thank L. Blanchet, A. Buonanno, and S. Detweiler for a careful reading of the manuscript and useful comments, as well as E. Poisson and C. M. Will for advice. The author acknowledges support from NSF through Grant No. PHY-0903631 and from the Maryland Center for Fundamental Physics.

\section*{References}

\bibliographystyle{iopart-num}
\bibliography{}

\end{document}